\documentclass[aps,prl,twocolumn,groupedaddress,showkeys]{revtex4-1}
\usepackage{graphicx}
\usepackage{dcolumn}
\usepackage{bm}
\usepackage{citesort}

\def \const {{\rm const}}

\def \be {\begin{equation}}
\def \ee {\end{equation}}
\def \ba {\begin{eqnarray}}
\def \ea {\end{eqnarray}}

\begin{document}

\title{Fluid flow control with transformation media}

\author{Yaroslav A. Urzhumov and David R. Smith}
\affiliation{Center for Metamaterials and Integrated Plasmonics, Pratt School of Engineering, Duke University, Durham, North Carolina 27708, USA}
\email{yaroslav.urzhumov@duke.edu}

\date{\today}

\begin{abstract}
We introduce a new concept for the manipulation of fluid flow around three-dimensional bodies. Inspired by transformation optics, the concept is based on a mathematical idea of coordinate transformations and physically implemented with anisotropic porous media permeable to the flow of fluids. In two situations --- for an impermeable object placed either in a free-flowing fluid or in a fluid-filled porous medium --- we show that the object can be coated with an inhomogeneous, anisotropic permeable medium, such as to preserve the flow that would have existed in the absence of the object. The proposed fluid flow cloak eliminates downstream wake and compensates viscous drag, hinting us at the possibility of novel propulsion techniques.
\end{abstract}


\maketitle


Coordinate transformations, specifically, conformal maps, are widely used in fluid dynamics as a method of solving the biharmonic equation describing incompressible Stokes flow of viscous fluids~\cite{landau_lifshitz_fluid}, in domains with complicated boundary shapes. They are also applied~\cite{bear72} to solving the Laplace equation describing fluid pressure distributions in porous media (Darcy's law), and the Poisson equation of electrostatics.
As a subset of the separation of variables method, conformal transformations thus offer a unique and powerful approach to the {\it forward} problems of incompressible flow, as well as electro- and magnetostatics.

The utility of more general coordinate transformations in the {\it inverse} electromagnetic~\cite{pendry_smith06,cai_milton07,urzhumov_smith_prl10} and acoustic~\cite{cummer_njp07,norris_prs08,urzhumov_smith_njp10,zhang_fang_prl11} problems has been demonstrated recently, most impressively by showing the possibility of electromagnetic invisibility, dubbed {\it cloaking}~\cite{pendry_smith06,schurig_smith06,leonhardt_science06,cai_milton07,urzhumov_pendry_jopt11}.
The key idea that enabled the progress in those areas was the combination of coordinate transformations with coordinate-dependent, and often extremely exotic~\cite{pendry_ring99,smith_apl00,fang_zhang_natmat06,urzhumov_smith_prl10}, material properties. Transformation optics~\cite{pendry_smith06,leonhardt_science06,cai_milton07,chen_sheng_10,urzhumov_smith_prl10} and transformation acoustics~\cite{cummer_njp07,norris_prs08,urzhumov_smith_njp10} offer solutions to some inverse scattering problems~\cite{greenleaf_phys03} by reducing them to an elementary one -- for example, scattering off a point object in free space.

There is no reason why this conceptual approach cannot be used in other areas of physics~\cite{chang_hu10}, and, in fact, it has already been applied to conductive heat transfer~\cite{li_huang10}, linear elastodynamics~\cite{milton_willis_njp06,farhat_prl09,brun_apl09,scandrett_howarth10,nemat_nasser10,urzhumov_smith_njp10,urzhumov_smith_prl10}, surface wave~\cite{farhat_prl08} and quantum-mechanical matter wave~\cite{zhang_zhang_prl100} dynamics. The fundamental requirement for the applicability of this concept is the presence of a medium with sufficiently flexible properties, which enables manipulation of the coefficients in the equations describing the dynamical process. Here, we propose porous media as a substrate for {\it transformation fluid dynamics}, and investigate the feasibility of controlling certain features of incompressible fluid flow.

The electromagnetic cloak of invisibility, which inspired this approach, manipulates the field lines of electric and magnetic fields, and the streamlines of momentum flux, in such a manner that these lines are virtually indistinguishable from those in the absence of any object~\cite{pendry_smith06}. Because the streamlines are unperturbed in the free space outside the cloak, electromagnetic radiation avoids scattering off the structure and therefore exerts no electromagnetic pressure on it. Achieving a similar level of control over the streamlines of the fluid flow outside the structure would imply canceling the viscous drag exerted on the structure. In this paper, we demonstrate the possibility of fluid flow cloaking using porous media whose properties are more general than those considered in earlier studies~\cite{koplik_levine83,masliyah_ven87}.

\begin{figure}
\centering
\begin{tabular}{cc}
\includegraphics[width=0.5\columnwidth]{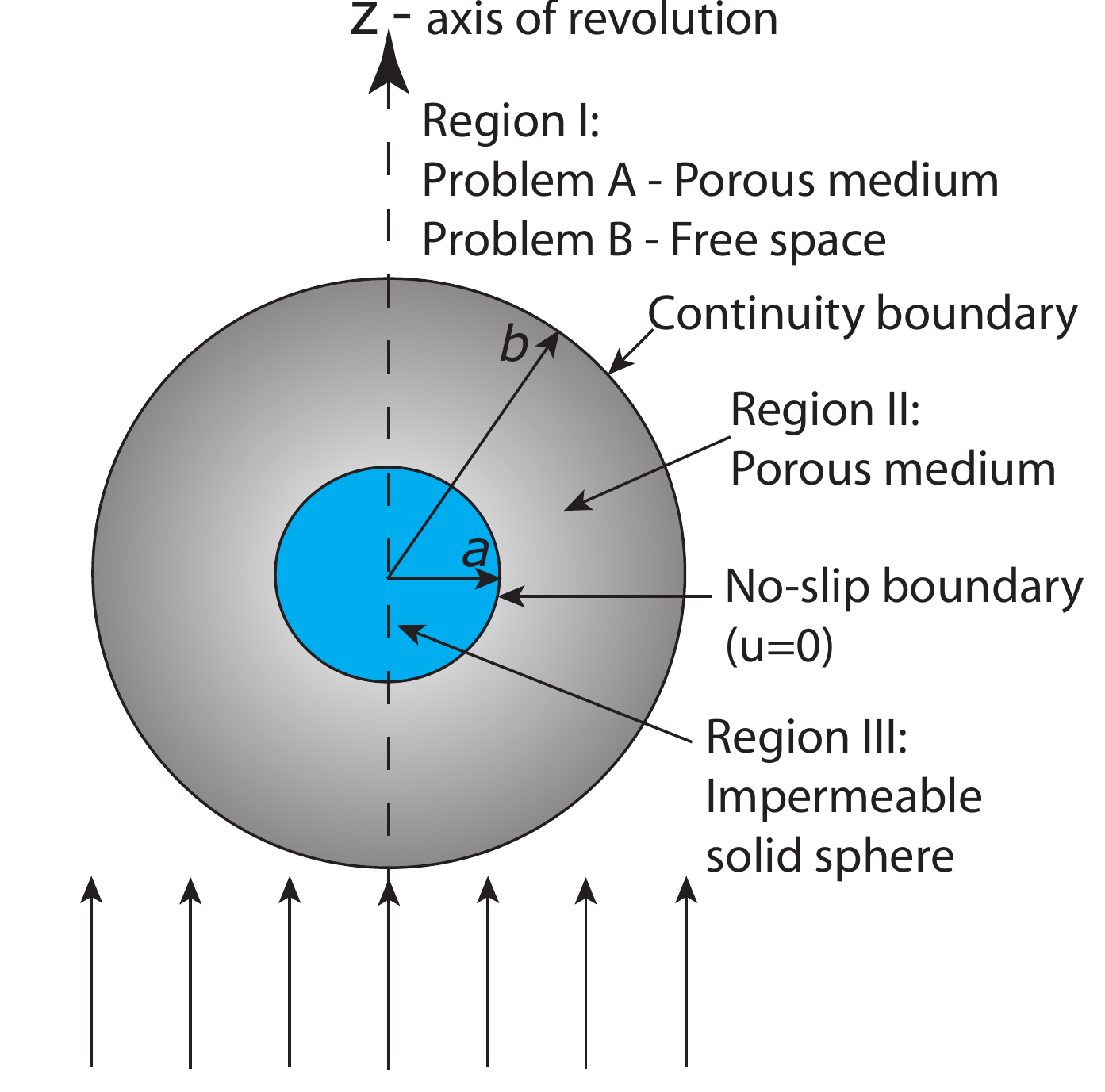}&
\includegraphics[width=0.4\columnwidth]{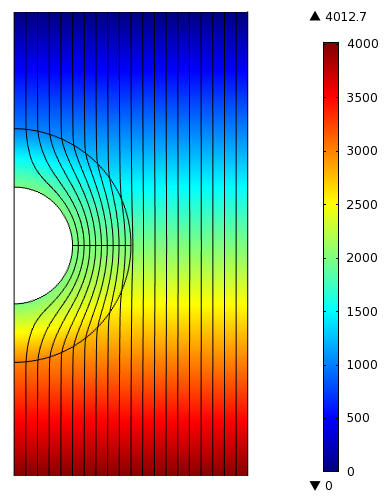}\\
(a)&(b)\\
\end{tabular}
\caption{(color online). (a) Schematic of the proposed fluid flow cloaking structure, and the definition of Problems A and B.
(b) Solution to Problem A: pressure distribution and streamlines of flow, calculated using the Brinkman-Stokes equation (\ref{eq:Brinkman-Stokes}) with $\kappa_0/b^2=10^{-3}$ in Region I. The solution is virtually indistinguishable from the one obtained with Darcy's law (not shown).
}
\label{fig:schematic_darcy}
\end{figure}


To allow a simple analytical treatment, we consider a solid spherical object of radius $a$ impermeable to the fluid, surrounded by a concentric permeable porous shell of exterior radius $b$, immersed in a fluid occupying an unbounded region (which we refer to as {\it free space}), and subject to a stationary, uniform flow with asymptotic velocity $u_0$ in the $z$-direction (Fig.~\ref{fig:schematic_darcy}(a)). For further simplification, consider the flow with a small Reynolds number, $Re = \rho_0 u_0 b/ \mu_0 \ll 1$, where $\rho_0$ and $\mu_0$ are the density and the dynamic viscosity of the fluid. In this regime, the nonlinear inertial term $(\vec u\cdot \vec \nabla)\vec u$ in the Navier-Stokes equations can be neglected, and the stationary incompressible flow is described by the Stokes equation:
\be
\mu_0 \nabla^2 \vec u =  \vec \nabla p.
\label{eq:Stokes}
\ee
The fluid velocity field $\vec u$ is divergence-free,
\be \vec \nabla \cdot \vec u=0.
\label{eq:continuity}
\ee
In the porous domain, the flow is described by the Brinkman-Stokes equation~\cite{bear72}:
\be
\mu \nabla^2 \vec u = \vec \nabla p + \mu_0 \kappa^{-1} \vec u,
\label{eq:Brinkman-Stokes}
\ee
where $\mu$ is the {\it effective viscosity} of the flow in the porous medium, and $\kappa$ is called the {\it permeability}.

In general, for an anisotropic porous medium, $\kappa^{-1}$ is a rank-two tensor~\cite{bear72}, and $\mu$ is a tensor of rank four. While the implementation of anisotropic permeability using layered media is well-known (see \textsection 5.8.3, ``A Layered Medium as an Equivalent Anisotropic Medium'' in Ref.~\cite{bear72}), effective viscosity of fluids in porous media has been a subject of some debate~\cite{masliyah_ven87}. To evaluate the degree of control that can be achieved by manipulating only permeability, in this study we assume that the effective viscosity in the porous medium ($\mu$) is isotropic, constant, and equal to the fluid viscosity in the absence of a porous medium ($\mu_0$); this assumption is consistent with established practice~\cite{masliyah_ven87}.
Whenever a numerical calculation occurs, we assume for concreteness that $\rho_0=10^3$~kg/m$^3$ and $\mu_0=10^{-3}$~Pa$\cdot$s, which approximates water at room temperature. Additionally, it is assumed that the radius of the structure $b=1$~mm.


In the limit of small permeability, $\kappa \ll b^2$, the term $\mu \nabla^2 \vec u$ in Eq.~(\ref{eq:Brinkman-Stokes}) can be neglected in comparison with the term $\mu_0 \kappa^{-1} \vec u$, and the Brinkman-Stokes equation reduces to Darcy's law~\cite{bear72},
\be
 \vec \nabla p  = - \mu_0 \kappa^{-1} \vec u,
\label{eq:Darcy_law}
\ee
which, combined with the continuity equation for an incompressible fluid (\ref{eq:continuity}),
gives Darcy's pressure equation~\cite{bear72}:
\be
 - \vec \nabla \kappa \vec \nabla p = 0.
\label{eq:Darcy_Laplace}
\ee

If we allow the porous medium to fill the entire space around the solid sphere, Darcy's equation (\ref{eq:Darcy_Laplace}) is at least approximately valid in all regions of the flow. We then formulate {\bf Problem A} as follows:
assuming that the porous medium in Region I ($r>b$) is uniform and isotropic with permeability $\kappa_0=\const$, and that the permeability tensor $\kappa$ is uniaxially-anisotropic and radially-symmetric in Region II ($a<r<b$), i.e.
\be
\kappa=\kappa_r(r)\hat r\hat r + \kappa_t(r) \hat \theta \hat\theta + \kappa_t(r) \hat\phi \hat\phi,
\label{eq:kappa_tensor}
\ee
find the functions $\kappa_r(r)$ and $\kappa_t(r)$
such that a solution exists in which the velocity is constant everywhere in Region I:
\be
\vec u(r,\theta)= u_0 \hat z = u_0 (\hat r\cos\theta - \hat\theta \sin\theta).
\label{eq:uniform_flow}
\ee
Note that because Region I contains a porous medium, there is a pressure gradient in the $z$-direction even for a uniform (``plug") flow (\ref{eq:uniform_flow}), in which the viscous fluid stress is absent.

The formulation of the more practically important {\bf Problem B} is the same as that of Problem A, except that Region I ($r>b$) contains no porous medium. In both Problems, we require the standard, no-slip boundary condition, that is, $\vec u=0$ at $r=a$, which corresponds to a surface of an impermeable solid (Region III).

Because Problem A in Darcy's approximation is mathematically equivalent to the problem of cloaking a spherical object from a uniform electrostatic field (except for the boundary condition), we could simply use the solution obtained in Ref.~\cite{pendry_smith06},
\be
\kappa_r(r)=\kappa_0\frac{b}{b-a}\left(\frac{r-a}{r}\right)^2,\,
\kappa_t(r)=\kappa_0\frac{b}{b-a},
\label{eq:problemA_solution}
\ee
which corresponds to the following transformation:
\ba
r'=b(r-a)/(b-a), \theta'=\theta, \phi'=\phi.
\label{eq:linear_transform}
\ea
While the boundary condition on the surface of the cloaked object ($r=a$) used in electrostatics is different from the no-slip boundary condition for the fluid flow, in three dimensions a spherical cloak described by Eq.~(\ref{eq:problemA_solution}) remains a perfect cloak regardless of the kind of homogeneous boundary condition used at $r=a$; see Ref.~\cite{urzhumov_pendry_jopt11} for a discussion of the boundary conditions in two and three dimensions.

The pressure distribution and the streamlines of plug flow around the structure (\ref{eq:problemA_solution}) obtained in the approximation of Darcy's law are shown in Fig.~\ref{fig:schematic_darcy}(b), proving that the distributions (\ref{eq:problemA_solution}) represent an exact solution to Problem A.  The plotted solution was calculated through the axisymmetric formulation of the Brinkman-Stokes equation available as the porous flow model in the finite element analysis package COMSOL Multiphysics~\cite{comsol10}.
Anisotropic permeability in the Brinkman model was implemented by customizing the stress tensor prescribed by the standard model.
When the permeability $\kappa_0$ is small enough, the viscous contribution to the fluid stress is negligible compared to the Darcy's force, and the solution of Brinkman-Stokes equation is indistinguishable from the solution of the Darcy's equation.


Now, we turn to the more interesting Problem B where the unbounded Region I is free of porous medium (see Fig.~\ref{fig:schematic_darcy}(a)), and the flow is described by the Stokes equation in Region I and the Brinkman-Stokes equation in Region II. The particular case of a creeping flow around a uniform, isotropic porous shell has been studied by various authors using the Stokes stream function~\cite{koplik_levine83,masliyah_ven87} and other methods~\cite{vasin_starov08,tsai08}. Here, we generalize the known solutions~\cite{koplik_levine83,masliyah_ven87} of the forward problem to allow the inclusion of inhomogeneous, although still radially-symmetric, uniaxially anisotropic permeability (\ref{eq:kappa_tensor}), and use them to solve the inverse problem (Problem B).


In Region I (absent of any porous medium), the most general solution to Eq.~(\ref{eq:Brinkman-Stokes}) that satisfies the condition of plug flow
(\ref{eq:uniform_flow}) at $r\to\infty$ can be written as~\cite{masliyah_ven87}
\ba
u_r
= u_0 \cos\theta  \left(1 - \frac{A}{r^3} - \frac{B}{r} \right), \nonumber \\
u_\theta
= -u_0\sin\theta  \left(1 + \frac{A}{2r^3} - \frac{B}{2r} \right),
\label{eq:free_space_flow}
\ea
where $A$ and $B$ are some coefficients that depend on the details of the flow in the porous Region II. In the exact cloaking solution we are seeking, both $A=B=0$; this means that all components of the viscous stress tensor, $\sigma^{visc}_{ij}=\mu_0(\partial_i u_j+\partial_j u_i)$, as well as the hydrostatic pressure $p$ (gauged to $p=0$ at infinity) vanish everywhere in Region I, leading to zero fluid stress $\sigma_{ij}=-p\delta_{ij}+\sigma^{visc}_{ij}$ and a vanishing drag force. The latter can be calculated as
$F_z=\oint_{S_R} \sigma_{zj} dS_j$, where $S_R$ is a sphere of radius $R$. It is easy to see that in the limit $R\to \infty$ only the terms $\sim B/r$ in Eq.~(\ref{eq:free_space_flow}) give finite contributions. Thus, the drag force is proportional to the coefficient $B$ and is entirely independent of $A$~\cite{masliyah_ven87}. Therefore, for the purpose of eliminating only the drag force, but not the laminar wake (deviation of the flow from uniformity), a weaker form of the cloaking problem can be postulated that only requires $B=0$ without a constraint on $A$. This wider class of ``weak cloaking'' solutions is not analyzed in this work.

\begin{figure}
\centering
\begin{tabular}{ccc}
\includegraphics[height=0.5\columnwidth]{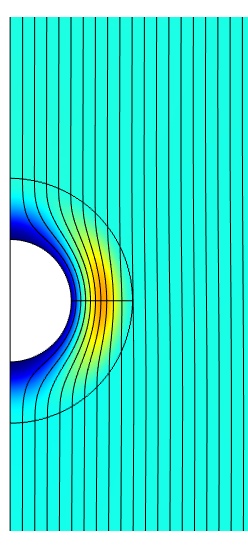}&
\includegraphics[height=0.5\columnwidth]{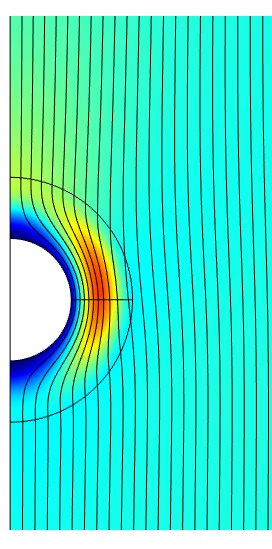}&
\includegraphics[height=0.5\columnwidth]{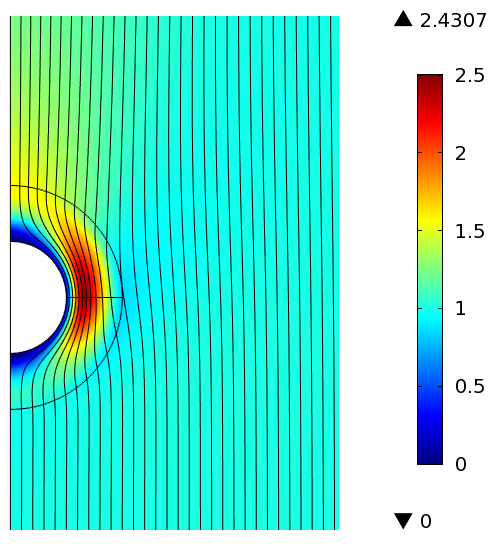}\\
(a)&(b)&(c)\\
\end{tabular}
\caption{(color online). Solutions to Problem B: maps of the axial velocity normalized to velocity at infinity ($u_z/u_0$) and streamlines of flow around and through the porous spherical shell. Left to right: $Re\le 0.5$ (a), $Re=2.5$ (b) and $Re=4.5$ (c).
}
\label{fig:brinkman_stokes_cloak}
\end{figure}

To find the exact cloaking solution ($A=B=0$), we employ numerical optimization of the Brinkman-Stokes equation (\ref{eq:Brinkman-Stokes}) with the unknown anisotropic permeability profile using COMSOL~\cite{comsol10}.
The inertia terms in both the Brinkman and free-space Navier-Stokes equations were neglected.
The optimization solver discretizes the unknown one-dimensional functions of the spherical radius, $\kappa_{r}^{-1}(r)$ and $\kappa_t^{-1}(r)$, and treats the nodal values of these functions as the optimization variables to solve for. The optimization goal minimized by the solver is chosen as $A^2+B^2$; the fully converged solution has $A \approx B \approx 0$ with a numerical accuracy of better than $10^{-8}$. Consequently, the flow pattern everywhere outside the porous shell satisfies the requirement $\vec u=\const$, as seen from Fig.~\ref{fig:brinkman_stokes_cloak}(a).

\begin{figure}
\centering
\begin{tabular}{c}
\includegraphics[width=0.8\columnwidth]{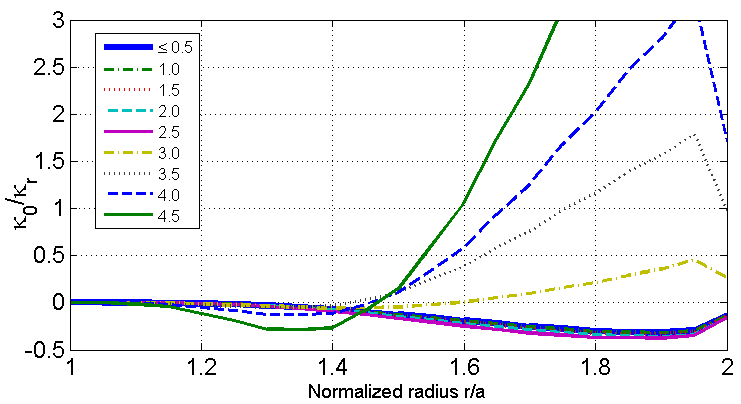}\\
\includegraphics[width=0.8\columnwidth]{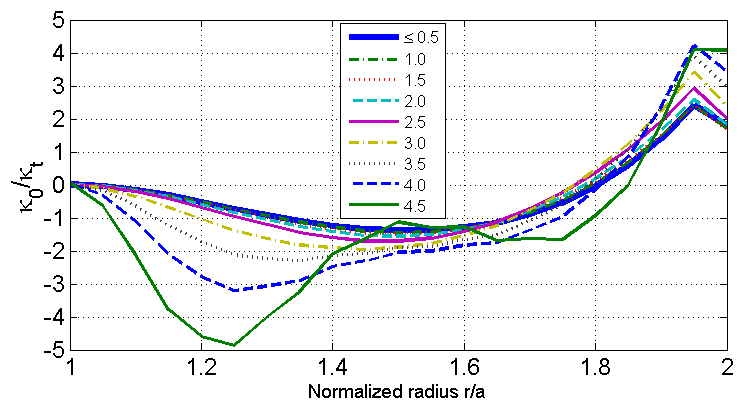}
\end{tabular}
\caption{(color online). Solutions to Problem B:
radial (top) and azimuthal (bottom) components of inverse permeability (normalized by $\kappa_0=a^2/4$) as functions of the spherical radius $r$
corresponding to solutions of the exact cloaking problem ($A=B=0$) with various Reynolds numbers.
}
\label{fig:permeability_profiles}
\end{figure}

The distribution of $\kappa$ components necessary to implement this flow pattern using a shell with the aspect ratio $b/a=2$ is shown as the thick solid line in Fig.~\ref{fig:permeability_profiles}.
The function $\kappa_r(r)$ is negative in its entire range $a<r<b$, and the function $\kappa_t(r)$ is only positive for $r_0<r<b$ and negative for $a<r<r_0$, where $r_0\approx 1.8a$.
The medium with a negative permeability is an active medium: it must use external energy to provide acceleration rather than deceleration to the fluid permeating through the porous medium. A physical implementation of such a medium can be foreseen as a distribution of minute pumps that help propel the fluid and compensate in a controlled fashion the viscous pressure drop that exists due to the velocity gradient in the $a<r<b$ region. Potential candidate technologies for such a directional acceleration on the micro-scale level are electro-osmotic~\cite{chen_chang05}, pneumatic and piezoelectric~\cite{lintel_bouwstra88} micro-pumps.

So far we have been neglecting the nonlinear advective term in the Brinkman equation, which is only valid for Reynolds numbers $Re\ll 1$. In this regime, the permeability profile is independent of the asymptotic velocity $u_0$. Using numerical optimization of the full Brinkman and Navier-Stokes equations,
\be
\rho_0 (\vec u \cdot \vec \nabla)\vec u = -\vec \nabla p + \mu_0 \nabla^2 \vec u - \mu_0 \kappa^{-1} \vec u,
\label{eq:Brinkman-Stokes-nonlinear}
\ee
where $\kappa^{-1}\equiv 0$ in Region I, we have also investigated the possibility of higher-$Re$ flow cloaking.
In Eq.~(\ref{eq:Brinkman-Stokes-nonlinear}), we assume that porosity in Region II, $\epsilon_p$, equals unity, which is consistent with our assumption $\mu=\mu_0$.

Figures~\ref{fig:brinkman_stokes_cloak}(a-c) demonstrate that the flow can be cloaked for Reynolds numbers up to approximately 4.5; at larger $Re$ the assumption that the flow in Region I is given by Eq.~(\ref{eq:free_space_flow}) is no longer accurate, and the optimization goal based on the $A$ and $B$ coefficients loses its meaning. The drag force $F_z$, normalized to the ideal Stokes drag for a sphere of radius $a$ ($F_{St}=6\pi\mu u_0 a$~\cite{landau_lifshitz_fluid}), was estimated by direct integration of the fluid stress, and shown to be $\le0.01$ at all $Re\le 1$, $\approx 0.157$ at $Re=2.5$ (Fig.~\ref{fig:brinkman_stokes_cloak}(b)), and $\approx 0.48$ at $Re=4.5$ (Fig.~\ref{fig:brinkman_stokes_cloak}(c)).
The resulting permeability profiles for various $Re$ values are shown in Fig.\ref{fig:permeability_profiles}. We observe that these profiles are virtually independent of the velocity magnitude up to $Re \approx 2.5$, but above that value they evolve quickly with $Re$.

In conclusion, we have postulated and solved the problem of cloaking a spherical object from the fluid flow
in a uniform porous medium (A), and in free space (B), using a permeable porous shell. The new concept of utilizing permeable media for fluid flow management opens the door to novel hydrodynamical approaches to compensating the viscous drag force, eliminating laminar wake and inhibiting the formation of turbulent wake past moving bodies. The emergence of these strategies hints at the possibility of wake-free distributed propulsion systems.


This work was supported by the U.S. Navy through a subcontract with SensorMetrix (Contract No. N68335-11-C-0011),
and partially supported through a Multidisciplinary University Research Initiative, sponsored by the U.S. Army Research Office (Grant No. W911NF-09-1-0539).

\bibliographystyle{unsrt}
\bibliography{CFD_cloak}

\end{document}